\documentclass[11pt,twoside]{article}
\usepackage{jltp_latex2e}

\newcommand{\be}{\begin{equation}}
\newcommand{\ee}{\end{equation}}
\usepackage[dvips]{graphicx}

\title{Step Structure Observed in the Superconducting Transition of a Bulk Pb Wire with Very Low Current Densities}

\author{X. Jehl, D. Braithwaite, P. Payet-Burin and R. Calemczuk}

\address{Service de Physique Statistique, Magn\'etisme et Supraconductivit\'e, 
\\DRFMC, CEA-Grenoble, 38054 Grenoble Cedex 9, France}

\runninghead{X. Jehl {\it et al.}}{Step structure in the superconducting transition of bulk lead}

\begin{document}

\begin{abstract}
We have developped a SQUID picovoltmeter and precise temperature control ($\sim \mu K$) to measure sharp $R(T)$ superconducting transitions of low impedance samples, with low current densities. We present measurements obtained on a bulk cylinder of lead ($\phi 0.25mm$) in transverse magnetic field. A distinctive hysteretic step structure is observed. It is washed out on increasing the current, and reinforced in higher field. Classical models for the dissipation mechanisms in the intermediate state are briefly reviewed and used to give a first interpretation of these results, although a complete understanding is not reached yet.


\end{abstract}

\maketitle

\section{INTRODUCTION}

\par
\begin{figure}[h]
\begin{center}
\includegraphics[width=11cm]{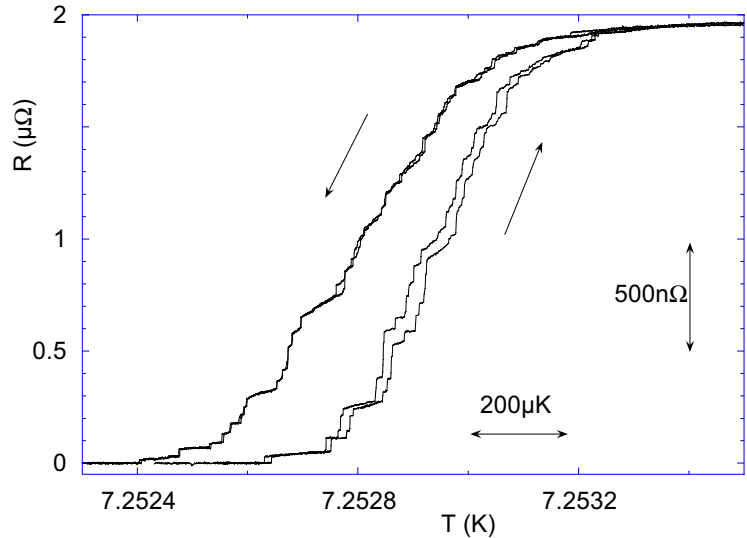}
\caption{Step structure observed in the third sample $Pb3$ with DC current $380\mu A$ in earth's field.}
\label{fig:pb3}
\end{center}
\end{figure}

 For fundamental and technological reasons the study of dissipation mechanisms in superconductors has always been very important. In the early $60s$ the discovery of type II superconductivity increased the interest of the field, and a number of studies concerning the magnetic flux structures in the intermediate or mixed state were performed. $V(I)$ curves \cite{lalevic, warburton, newborer, meyer} and noise measurements \cite{vangurp} gave major results combined with purposely developped experimental techniques such as direct visualization of field distribution \cite{kampwirth,parks} and the dc-flux transformer \cite{solomon}. This resulted in a comprehensive knowledge of the dissipation regimes due to magnetic flux penetration in both type I and type II superconductors \cite{huebener}.
\\Step structures have been observed in $V(I)$ curves of thin films or filaments \cite{meyer} and were attributed to phase slip \cite{tinkhamsbt}, the keypoint of which being the size of the sample smaller than the coherence length $\xi (T)$.
\\Voltage steps have also been seen in current induced transitions of 3D strips and were interpreted in terms of the dynamics of flux tube arrays \cite{huebener2,chimenti}.
\\In all cases the transition was induced by high currents, and current was known to play an important role as it interacts with the field and therefore with the magnetic flux structures \cite{parks}. As critical currents in superconductors decay to zero when approaching $Tc$, one can expect similar results when keeping the current constant and increasing the temperature. Although some structures in R(T) transitions have been seen \cite{warburton}, very few studies seem to have been published, perhaps for technical reasons. It is important to have a temperature resolution much smaller than the transition width of the pure sample ($\sim mK$) as well as to have a very sensitive resistance measurement. Both conditions can be fulfilled using modern instrumentation. 
\\We performed very sensitive $R(T)$ measurements on a bulk lead cylinder using a SQUID based picovoltmeter and a thermal system allowing precise measurement within very sharp transitions. The original results we present consist in a highly reproducible step structure with hysteresis [Fig.\ref{fig:pb3},\ref{fig:pb3zoom}]. The effect of current and local field are also presented, and we discuss possible interpretations in view of existing models of dissipation mechanisms in the intermediate state.

\section{EXPERIMENTAL}
\begin{figure}[h]
\begin{center}
\includegraphics[width=8cm]{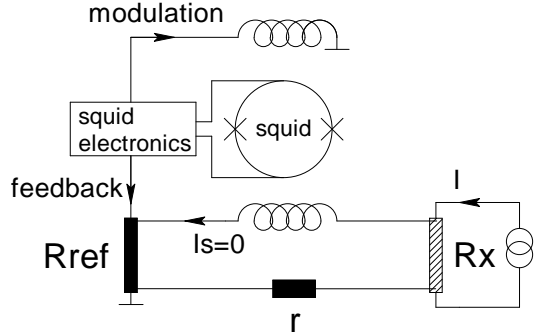}
\caption{Schematic of the circuit. The SQUID input coil measures the balance of the bridge composed by the sample $Rx$ and reference $Rref$. Contact resistances are represented by r which determines the sensitivity as $r \gg Rref,Rx$. Typical values are $r=2.4m\Omega$, $Rref=67\mu\Omega$ and $Rx: 0$ to a few $\mu\Omega$.}
\label{fig:circuit}
\end{center}
\end{figure}

After pionnering works by Clarke \cite{clarke} , picovoltmeters based on SQUIDs have been developped to measure small electrical signals \cite{gallop,sachslehner,krasnopolin,barbanera}. In this work we use the SQUID to detect small currents ($nA$) through small resistances ($m\Omega$). We obtain a resolution of the order of one picovolt, i.e. a few $n\Omega$ for measuring currents of about $400\mu A$. The general principle of resistance measurement is a bridge composed with the sample resistance $Rx$ and a reference resistance $Rref$, maintained at constant temperature, of well known value, both connected with 4 wires [Fig.\ref{fig:circuit}]. The input coil of the SQUID is connected in series between $Rx$ and $Rref$ so the SQUID is sensitive to the current in the voltage leads. Other unwanted resistances in series with the voltage leads (like contact resistances) are represented by $r$, which is much greater than $Rx$ and $Rref$ and therefore determines the sensitivity. The feedback current which is necessary to keep a constant flux on the SQUID is injected in the resistance bridge through $Rref$ as shown in Fig.[\ref{fig:circuit}]. Consequently the constant current $Is$ which flows in the voltage leads of the sample can be chosen to be zero.
\\We used a commercial Conductus DC-SQUID \cite{conductus} controlled by an electronic system designed and built in our laboratory. Double twisted shielded conductors and RF filters are carefully designed to reduce the noise on the SQUID. Performances and stability are enhanced with superconducting shields. A first global shield is made with a lead foil soldered around calorimeter in the helium bath. Then the whole voltage circuit connected to the SQUID, made of superconducting twisted wires, is inside tinned tubes. Thermalisation contacts and reference resistance (made of CuBe alloy) are in a closed lead box. This results in real white noise spectrum showing no peaks. RF suppression makes it possible to operate with DC currents as there are no more flux jumps. The SQUID and its electronis have characteristic current noise equivalent to $In=11{pA/\sqrt {Hz}}$  in the input coil. This is equivalent to the fundamental thermal noise of $R=2\Omega$ at $T=4.2K$, according to the Johnson-Nyquist relation:
\be
\label{1}
In=\sqrt{{4k_{B}T}\over{R}}
\ee
As the total impedance of the circuit ($\approx 2.4m\Omega$) is much lower, its thermal noise dominates. For $r=2.4m\Omega$ and $I=380\mu A$ (DC or AC(rms)) the current noise on $Is$ is $1{nA/\sqrt Hz}$ (DC) or $480{pA/\sqrt Hz}$  (AC, lock-in detection technique). On the sample this means  $2.4{pV/\sqrt Hz}$ (DC) or $1.1{pV/\sqrt Hz}$ (AC), and finally a resistance sensitivity of $6n\Omega$ (DC) or $3.4n\Omega$ (AC).  
\\ A Germanium thermometer with sensitivity $15mK/\Omega$ around $7K$ is read with a four leads AC resistance bridge with a current of $10\mu A$. It gives a resolution of $20\mu K$. Considerable improvement is made by smoothing the raw temperature data after very slow and regular temperature drifts, at speeds as low as $0.3mK/h$ over $5$ hours. The sample holder has a very large thermal time constant (heat capacity equivalent to $1.5Kg$ of copper at $7K$) and low thermal conductances to the heat sources. Therefore the system is always at equilibrium and the observed noise on temperature is not due to actual variations of the sample holder real temperature but only to the reading noise of the thermometry.
\\Three samples from a $0.25mm$ diameter polycristalline lead wire \cite{goodfellow} were successively prepared. In all cases four gold wires are connected with about $4mm$ between the voltage leads and $1cm$ total length. The first samples ($Pb1$ and $Pb2$) were measured with AC currents, the third sample ($Pb3$) with AC and DC currents. $Pb3$ had spotwelded contacts and showed a $1mK$ transition width and the lowest residual resistivity ($RRR=500$). The wire is horizontal in the earth's magnetic field, therefore local field near the wire is inclinated by $45^{o}$, with both axial and transversal component.

\section{RESULTS}  
\begin{figure}[h]
\begin{center}
\includegraphics[width=8cm]{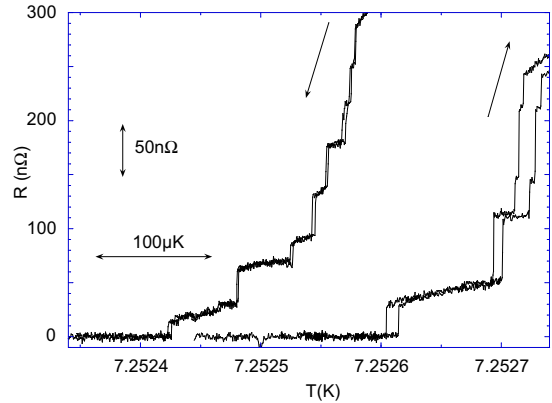}
\caption{Close up on the foot of the transition shown in Fig.[2].}
\label{fig:pb3zoom}
\end{center}
\end{figure}
\begin{figure}[h]
\begin{center}
\includegraphics[width=8cm]{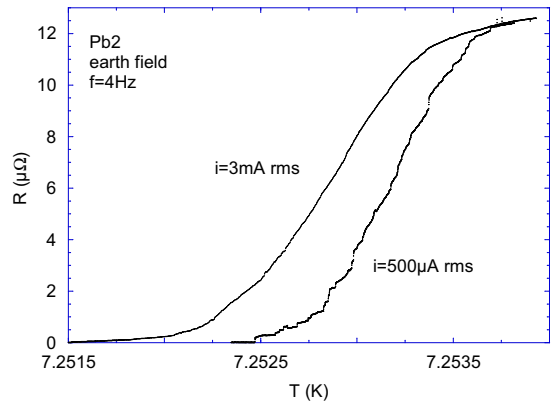}
\caption{Effect of current: the step structure is washed out on increasing slightly the current.}
\label{fig:courants}
\end{center}
\end{figure}

  
Fig.[\ref{fig:pb3}] shows a typical transition where the step structure and the hysteresis are visible. Similar results were found on all three samples. Fig.[\ref{fig:pb3zoom}] is a close up on the foot of the transition. Steps are different for increasing and decreasing temperature, almost perfectly reproducible in each case although some single steps are sometimes separated into sub-steps. Down-steps were also observed, and were reproducible too (Fig.[\ref{fig:2gauss}]). Fig.[\ref{fig:courants}] shows the effect of changing the current.
\\As current is increased from $0.5$ to $3mA$, $Tc$ is lowered, the transition is broadened and the step structure is radically changed, and almost disappears with $3mA$. Questions about the influence of AC currents on magnetic flux structures in the intermediate state motivated DC measurements on $Pb3$. Small changes were detected although the structure remained essentially the same. Indeed results with $380\mu A$ DC or AC(rms) showed slightly different steps, probably due to a peak current of $540\mu A$ in the AC case.

\begin{figure}[h]
\begin{center}
\includegraphics[width=10cm,height=6cm]{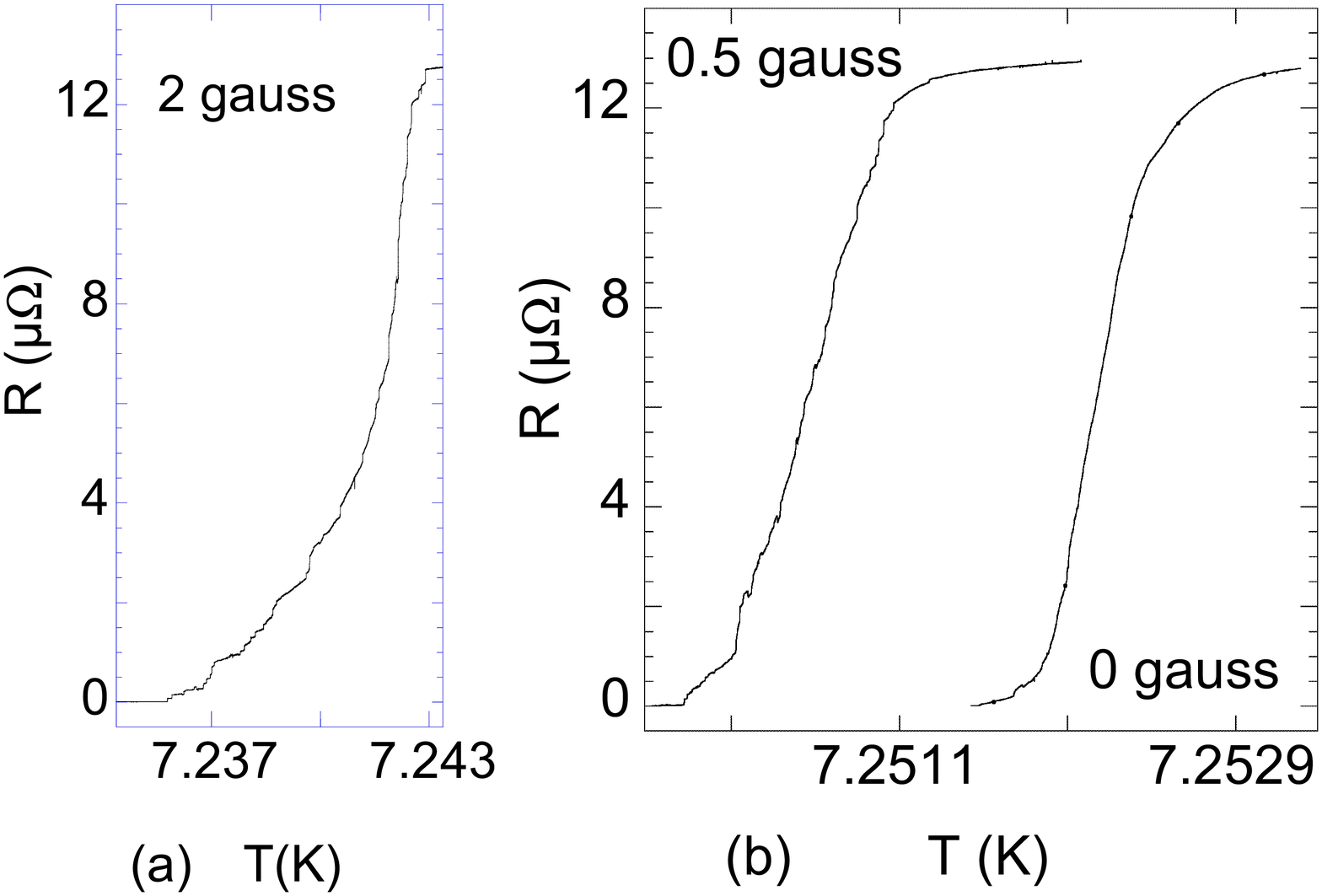}
\caption{Effect of external field seen on Pb2 with AC current $500\mu A$ rms at 4Hz:(a) the step structure is considerably reinforced under 2 Gauss. (b) the step structure is nearly destroyed under 0 Gauss. As expected $Tc$ and the transition width are also modified.}
\label{fig:fig5}
\end{center}
\end{figure}

\begin{figure}[h]
\begin{center}
\includegraphics[width=8cm]{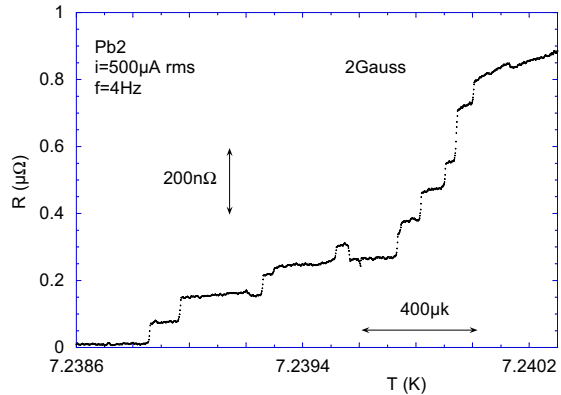}
\caption{Reinforcement of the step structure with increasing external field: close up on the foot of the transition under $2 gauss$. Note the down steps observed near $7.2396K$ and $7.2401K$ which are reproducible.}
\label{fig:2gauss}
\end{center}
\end{figure}

To study the effect of the external field we modified the earth's field first by compensating the vertical component and then applying $2G$ [Fig.\ref{fig:fig5}, \ref{fig:2gauss}] . A more extensive study of the external field effect was experimentally impossible at this time. The structure disappears when the transverse field is nearly zero. Under $2G$ the shift of $Tc$ and the broadening of the transition agree with $ {{dH_{c}}\over {dT}}\approx 0.25G/mK $ near $Tc$. The step structure is more pronounced as field increases [Fig.\ref{fig:2gauss}].
 We also performed $V(I)$ measurements within the transition which revealed a very ohmic behaviour and not a step structure. The low currents we used (max. $1mA$) create a self field lower than $0.05G$ near the surface of the wire. Therefore (as long as it is not compensated) the external field ($0.5G$ and $2G$) is also the local field. Increasing the current (and therefore the field produced by this current) tends to wash out the structure. Reducing the external field at constant current has the same effect. Therefore it seems we only observe the step structure when the transition is caused by the external field and not the current. The very low current densities necessary to observe the structure are perhaps the main reason why it has not been observed before. 

\section{DISCUSSION}

It is well known that a bulk type I superconductor with demagnetizing factor $D$ is in the intermediate state in the magnetic field range mbox{$Hc(1-D) < H < Hc$.} In our geometry ($D={1\over2}, H{_\perp}=0.5G$) the intermediate state should take place over the range of field between $0.25G$ and $0.5G$, i.e. on $1.1mK$. This is in good agreement with our results [Fig.\ref{fig:pb3}]. The question of whether dissipation should appear as soon as normal domains are nucleated within the sample (when $H={1\over2}Hc$) or later has been central in early works on the intermediate state \cite{london,andrew,shoenberg}. The prediction that dissipation should actually start at ${1\over 2}Hc$ as $T\rightarrow Tc$ is well confirmed by our results.     
\par The intermediate state is composed with normal ($N$) and superconducting ($S$) domains. While magnetization measurements can be interpreted by considering the macroscopic flux which enters the sample, resistance shall clearly be sensitive to the exact arrangement of the domains. The structure is determined by a compromise between the negative energy gained through relaxation of the field distortion and the positive energy paid in establishing super-normal boundaries. The two relevant domain shapes are threads or laminae perpendicular  to the cylinder axis and parallel to the transverse field. The two mechanisms responsible for dissipation are the ohmic resistance of normal regions across the sample and the flux flow dissipation mechanism due to the dynamics of normal regions \cite{kim}. Although they are generally difficult to separate, it has been shown by noise measurements \cite{vangurp} and dc flux transformer technique \cite{solomon} that at low current and low field the flux flow mechanism dominates while near $Hc$ (at the end of the transition) the ohmic regime dominates. 
Hysteresis can be understood as a combination of supercooling and superheating \cite{burger}. Near $Tc$, the  Landau-Ginzburg $\kappa$ value of pure lead is such that $Hc_3<Hc$ \cite{cardona}. This can give rise to a supercooling effect. Impurities or surface defects usually limit this effect as they act as nucleation centers. However near $Tc$ they are not efficient and the supercooling can approach that of the ideal metal \cite{faber}. The surface barrier against flux penetration is responsible for superheating \cite{bean}. The measured hyteresis is compatible with predictions \cite{cardona,burger}.
The mechanism of phase slip \cite{tinkhamsbt} can not be responsible for the step structure because the coherence length $\xi$ is small compared to the sample size when steps are first observed. At $1mK$ under $Tc$, $\xi$ calculated in the clean limit ($\propto {(1-{({T\over{Tc}})^{2}})^{-1\over{2}}}$) is $50\mu m$, and steps are observed several $mK$ under $Tc$ under 2 Gauss.  
\par Voltage steps were observed by Huebener and coworkers in the current-induced resistive state of indium and lead strips\cite{huebener2}. The self field produced by strong currents generates flux tubes which move rapidly to the center where they annihilate with flux tubes of opposite signs nucleated at the opposite edges of the strip. The isolation of a single channel of flux trains at a constriction confirmed this mechanism \cite{chimenti}. The sudden appearance of this process at one point along the sample generates a voltage step in the $V(I)$ curve and a peak in noise measurements \cite{huebener2}. The nucleation rate related to the Josephson frequency  was also experimentally verified \cite{orlowski}.
\\However a second dissipation mechanism was revealed by instabilities in the current-induced resistive state\cite{lalevic} and noise power measurements\cite{huebener2}. In this case dissipation is due to the abrupt rearrangements of the distribution of currents and magnetic fields within the specimen.

 In a planar geometry the non uniformity of the field within the sample leads to geometrically metastable states \cite{waysand}. This is not the case in a long cylinder where demagnetizing effects result in a homogeneous internal field \cite{landau}. Therefore the flux can be located wherever inside the specimen as it is energetically equally favorable. 
 Also because of its Ginzburg-Landau $\kappa$ parameter close to $1\over{\sqrt2}$, the positive surface energy term 
\be
{Hc^{2}\over{8\pi}}(\xi-\lambda)
\ee
is small in lead and the differences of energies between different configurations can also be very small. The overall structure of the intermediate state can therefore strongly fluctuate, giving rise to instabilities between metastable resistance levels and even to negative differential resistivity \cite{huebener3}. We observed these effects under the form of subdivision of steps into sub-steps (Fig.[\ref{fig:pb3}]) and reproducible down steps (Fig.[\ref{fig:2gauss}]).
We have not found any systematic relation defining the step sizes or the slope of the resistivity between them.  Very flat stages are observed far from the beginning of the transition while quite steep stages are present close from the ends. 
It is not clear why we don't observe the step structure when the transition is induced by the current whereas it was seen with very high current densities by Huebener. One difference is that in this case for the same local field, and therefore same macroscopic flux penetration, the force on the flux structure is proportionnal to the current, and therefore much stronger than when the transition is induced by an external field.

\section{CONCLUSION}

\par We have developped a SQUID picovoltmeter allowing the measurement of very low voltages ($\approx pV$) on low impedance samples, together with a very sensitive control of the temperature ($\approx \mu K$). We used this apparatus to investigate the resistive superconducting transition of a bulk type I lead wire ($R_N \approx \mu \Omega $) with low current densities ($\approx 1A/cm^{2}$) and in transverse magnetic field. The very sensitive $R(T)$ measurements we performed proved to be an original tool for investigating the intermediate state.
\par The hysteretic resistive transition (width $\approx 1mK$) is found to take place all over the range of the intermediate state, as theoretically expected. A reproducible step structure is observed all over the transition. At constant transverse applied field increasing slightly the current destroys the step structure. On the other hand increasing the field reinforces it considerably. 
\par The mechanism of flux penetration developped to explain the step structure observed in the current-induced resistive state of strips consists in the succession of trains of flux tubes rapidly moving to the center of the strip because of edge effects and symmetry of the self field induced by the current. 
In contrast with this case the local field induced by applying an external field has the same direction all around the sample and there is no geometrical opposition to flux motion in a cylinder. We believe these important differences are responsible for the disappearing of the step structure in increasing current and for its enhancement in higher field. 
We observe resistance jumps due to the sudden rearrangement of the domain patterns rather than nucleation and dynamics of flux tubes giving rise to flux-flow dissipation, as confirmed by the ohmic $V(I)$ curves.

\section*{ACKNOWLEDGMENTS}

We thank J.Guilchrist and H.Godfrin for helpful discussions.


\newpage
\begin{thebibliography}{99}


\bibitem{lalevic} B. Lalevic, {\it Phys.Rev.} {\bf 128}, 2, 1070 (1962) and B. Lalevic, {\it Phys.Stat.Sol.} {\bf 9}, 63, (1965).


\bibitem{warburton}  W.W. Webb and R.J. Warburton, {\it Phys.Rev.Lett.} {\bf 20}, 9, 461 (1968) and J.E. Lukens, R.J. Warburton and W.W. Webb, {\it Phys.Rev.Lett.} {\bf 25}, 17, 1180 (1970).

\bibitem{newborer} R.S. Newbower, M.R. Beasley and M. Tinkham, {\it Phys.Rev.B} {\bf 5}, 3, 864 (1972).

\bibitem{meyer} J. Meyer and G. Minnigerode, {\it Phys.Lett.} {\bf 38A}, 7, 529 (1972).


\bibitem{vangurp} G.J. Van Gurp and D.J. Van Ooijen, {\it Phys.Lett.} {\bf 17}, 230 (1965) and G.J. Van Gurp, {\it Phys.Lett.} {\bf 24A}, 10, 528 (1967).

\bibitem{kampwirth} R.P. Huebener, R.T. Kampwirth and V.A. Rowe, {\it Cryogenics} {\bf 12}, 100 (1972).

\bibitem{parks} J.D. Livingston and W. DeSorbo, {\it Superconductivity}, ed. R.D.Parks (Dekker), vol.2, chap.21 (1969).  

\bibitem{solomon} P.R. Solomon, {\it Phys.Rev.} {\bf 179}, 2, 475 (1969) and J.W. Ekin, B. Serin and J.R. Clem, {\it Phys.Rev.B} {\bf 9}, 3, 912 (1974).


\bibitem{huebener} R.P. Huebener, {\it Physics Reports} {\bf 13}, 4, 143 (1974) and R.P. Huebener, {\it Magnetic Flux Structures in Superconductors}, Springer-Verlag (1979) and R.P. Huebener {\it Rev.Mod.Phys.} {\bf 46}, 2, 409 (1974).



\bibitem{tinkhamsbt} W.J. Skocpol, M.R. Beasley and M. Tinkham, {\it J.Low Temp.Phys.} {\bf 16}, 1, 145 (1974) and M. Tinkham, {\it Introduction to superconductivity (2nd edition)}, McGraw-Hill, 427 (1996).


\bibitem{huebener2} R.P. Huebener and D.E. Gallus, {\it Phys.Rev.B} {\bf 7}, 9, 4089 (1973) and R.P. Huebener and H.L. Watson, {\it Phys.Rev.B} {\bf 9}, 9, 3725 (1974).  

\bibitem{chimenti} D.E. Chimenti and R.P. Huebener, {\it Sol.Stat.Comm.} {\bf 21}, 467 (1977).

\bibitem{clarke} J. Clarke, {\it Sci.Am.} {\bf 46} (1994) and J. Clarke, {\it proceedings of the IEEE} {\bf 77}, 8, 1208 (1989).

\bibitem{gallop} J.C. Gallop, {\it Squids, the Josephson Effects and Superconducting Electronics}, Adam Hilger, New York (1991).

\bibitem{sachslehner} F. Sachslehner and W. Vodel, {\it Cryogenics} {\bf 32}, 9, 805 (1992).

\bibitem{krasnopolin} I. Ya Krasnopolin, {\it Sov.Phys.JETP}, 1427 (1986). 

\bibitem{barbanera} S. Barbanera, M.G. Castalleno and V. Foglietti, {\it Rev.Sci.Instrum.} {\bf 59}, 7 (1988).

\bibitem{conductus} Conductus inc., {\it 969 West Maude Avenue, Sunnyvale CA 94086, USA}.

\bibitem{goodfellow} Goodfellow, {\it Cambridge Science Park, England CB4 4DJ}.

\bibitem{london} F.London, {\it Superfluids}, vol.1, Chap.C, Dover, New York (1961).

\bibitem{andrew} E.R. Andrew, {\it Proc.Roy.Soc.London} {\bf 194A}, 80 (1948).

\bibitem{shoenberg} D. Shoenberg, {\it Superconductivity}, Cambridge University Press, Cambridge (1952).

\bibitem{kim} Y.B. Kim, C.F. Hempstead and A.R. Strand, {\it Phys.Rev.} {\bf 131}, 2486 (1963) and Y.B. Kim, C.F. Hempstead and A.R. Strand, {\it Phys.Rev.} {\bf 139}, A1163, (1965).

\bibitem{burger} J.P. Burger and D. Saint-James, {\it Superconductivity}, ed. R.D.Parks (Dekker), vol.2, chap.16, 998 (1969). 


\bibitem{cardona} M. Cardona and B. Rosenblaum, {\it Low Temperature Physics LT9}, Plenum, New York, 560 (1965).

\bibitem{faber} T.E. Faber, {\it Proc.Roy.Soc.London} {\bf A241}, 531 (1957). 

\bibitem{bean} C.P. Bean and J.D. Livingston, {\it Phys.Rev.Lett.} {\bf 12}, 1 (1964).

\bibitem{orlowski} M.C.L. Orlowski, W. Buck and R.P. Huebener, {\it J. Low Temp. Phys.} {\bf 27}, 1, 159 (1977).

\bibitem{waysand} V. Jeudy, G. Jung, D. Limagne and G. Waysand, {\it Physica C} {\bf 225}, 331 (1994).

\bibitem{landau} L.D. Landau and L. Lifchitz, {Electrodynamics of Continuous Media}, Pergamon, ch.2, 42 (1960). 

\bibitem{huebener3} R.P. Huebener and D.E. Gallus, {\it Appl.Phys.Lett.} {\bf 22}, 11 (1973).




\end{thebibliography}
\end{document}